\documentstyle[12pt]{article}
\begin{document}

\baselineskip=0.4in

\rightline{SAGA-HE-107-96}

\rightline{July 1996}

\centerline{{\bf Quark condensate in nuclear matter 
 based on Nuclear Schwinger-Dyson formalism}} \par
\centerline{Tomohiro Mitsumori, Nobuo Noda ,\\ Hiroaki Kouno, Akira Hasegawa
 and Masahiro Nakano$^*$} \par

\centerline{\it Department of Physics, Saga University, Saga 840, Japan} \par
\centerline{\it $^*$University of Occupational and Environmental 
Health, Kitakyushu 807, Japan} 

\vskip 2cm

\noindent
{\bf abstract}

  The effects of
  higher order corrections of ring diagrams for the quark condensate are  
 studied by using the bare vertex
 Nuclear Schwinger Dyson  formalism
 based on $\sigma$-$\omega$ model.   
 At the high density the quark condensate is reduced by the higher 
 order contribution of ring diagrams more than the mean field theory 
  or  the Hartree-Fock .
\vfill\eject

The quantum chromodynamics (QCD) is a useful theory to describe quarks
 and gluons. Especially the perturbative QCD works very well at high
 energies. But at low energies the QCD shows non-perturbative behaviors.
  The QCD sum-rule approach is a useful tool in understanding the 
 property of hadrons in free space \cite{yazaki}. The vacuum expectation 
 value
 of the quark condensate is estimated by the sum-rule approach,
\begin{equation}
 <\bar qq>_0\simeq -(225\pm 25 {\rm MeV})^3 .
\label{a0} \end{equation}
 The QCD is not yet available 
 for hadronic phenomena , but there 
 have been several quantum hadrodynamics approachs to describe the properties
 of hadrons in nuclear matter. The quark condensate, which is shifted from
 vacuum values, is determined using these approachs .
 
Recently the quark condensate in medium has been evaluated in the 
 framework of mean field theory (MFT), 
 Nambu-Jona-Lasinio (NJL) model \cite{cohen1},
 and relativistic Brueckner Hartree-Fock (RBHF) \cite{li}\cite{brockmann} .
 The results of these methods show that
 the quark condensate in nuclear matter is reduced considerably 
 at the normal density.
 The Feynman-Hellmann  theorem
 relates the shift of quark condensate from vacuum value to the pion-nucleon
 sigma term $\sigma_{\pi N}$ which can be related to the 
 pion-nucleon scattering amplitude. 
 Using the Feynman-Hellmann theorem the quark condensate $<\bar qq>_\rho$ in nuclear
 matter is related to energy density $\epsilon$ as follows,

\begin{equation}
2m_q(<\bar qq>_\rho -<\bar qq>_0)=m_q{d\epsilon \over dm_q},
\label{a3}\end{equation}
where $<\bar qq>_\rho\equiv <\rho |\bar qq|\rho >$, and 
      $<\bar qq>_0\equiv <0|\bar qq|0>$. 
      The energy density $\epsilon$ in nuclear matter
can be written as      
$\epsilon=(m_N+E/A)\rho$ 
 ,where $m_N$ is the nucleon mass in free space and $E/A$ is the energy 
 per nucleon in nuclear matter and $\rho$ is the baryon density. 
If we neglect the dependence of meson-nucleon coupling constants
 on the current quark mass,  
 and using the Gell-Mann-Oakes-Renner relation,
 $2m_q<\bar qq>_0=-m_\pi^2f_\pi^2 $
(where $m_\pi \approx $138 MeV is the pion mass and $f_\pi \approx$
 93 MeV is the pion decay constant)
 and the definition of the pion-nucleon sigma term , $\sigma_{\pi N}
 =m_q (dm_N /dm_q)$ ,
 the quark condensate is obtained as following expression

\begin{equation}
{<\bar qq>_\rho \over <\bar qq>_0 }=1-{\rho \over m_\pi^2 f_\pi^2}
\Biggr [\sigma_{\pi N}+m_q {d \over dm_q}
\Biggr ( {E \over A} \Biggr )\Biggr ].
\label{b10}\end{equation}
\noindent
We accept $\sigma_{\pi N}=($45$\pm$ 7) MeV which was recently analyzed by 
 Gasser et al \cite{gasser}. 
 
 In this paper, we study the quark condensate in nuclear matter using
 the bare vertex Schwinger Dyson 
 (BNSD) formalism 
 \cite{nakano1}\cite{nakano2} based on $\sigma$-$\omega$ model and compare the
  result
 of the BNSD with the those of the MFT and the HF , since our purpose is 
 the examination of the contribution of the higher order ring diagrams
 for quark condensate. 
 Since we don't consider 
 the current quark mass dependence of coupling constants between 
 a nucleon and a meson,
 the derivatives of these 
 meson masses are approximately related to the nucleon mass \cite{saito}
 as follows,
\begin{equation}
\Biggr ({dm_i \over dm_q}\Biggr )/\Biggr({dm_N \over dm_q}\Biggr )
={m_i \over m_N},(i=\sigma\quad and\quad \omega).
\label{a8}\end{equation}

We acquire the final expression of the quark condensate in nuclear matter
 in $\sigma$-$\omega$ model , 
\begin{equation}
{<\bar qq>_\rho \over <\bar qq>_0 }=1-{\sigma_{\pi N}\over m_\pi^2 f_\pi^2}
\rho\Biggr [1+{\partial (E/A)\over \partial m_N}+\sum_{i=\sigma,\omega}
{\partial (E/A) \over \partial m_i}{m_i \over m_N}\Biggr ].
\label{a10}\end{equation}

 To evaluate the quark condensate in medium
 we need the energy density of the nuclear matter. But the direct results 
 from the QCD is not obtainable and so alternatively we calculate the energy 
 density in nuclear matter using the BNSD 
 in $\sigma$-$\omega$ model. This one modifies 
 the Fock exchange self-energy in the HF by replacing 
 free meson propagators with  full ones which are accompanied with 
 the particle-hole excitations.
We adopt the Walecka model\cite{walecka} which consists of three fields, the 
nucleon $\psi$, the scalar $\sigma$-meson $\phi$ and the vector 
$\omega$-meson
$V_\mu$ . The lagrangian density is given by
\begin{displaymath}
L = -\bar\psi(\gamma_\mu\partial_\mu +m_N)\psi -{1\over 2}(\partial_\mu 
\phi\partial_\mu\phi +m_s^2 \phi^2) 
\end{displaymath}
\begin{equation}
 -({1 \over 4}F_{\mu\nu}F_{\mu\nu}+{1 \over 2}m_v^2 V_\mu V_\mu)
+g_s\bar\psi\psi\phi +ig_v\bar\psi\gamma_\mu\psi V_\mu  , \label{b1} 
\end{equation}
\noindent
where $F_{\mu\nu}= \partial_\mu V_\nu - \partial_\nu V_\mu$ and $m_N$, $m_s$,
 $m_v$, $g_s$ and $g_v$ are nucleon mass, $\sigma$ meson mass, 
$\omega$ meson mass, $\sigma$-nucleon and $\omega$-nucleon 
coupling constants respectively. 
The nucleon propagator is obtained by following form,
\begin{equation}
G(k) = G^0(k)+G^0(k)\Sigma (k)G(k),\label{b2}
\end{equation}
\noindent
where $\Sigma$ is nucleon self-energy.
The mesons full propagators for the scalar and vector mesons 
are obtained 
 as follows \cite{nakano1} \cite{nakano2},
\begin{equation}
D_{ab}(k)=D_{ab}^0(k)+D_{ac}^0(k)\Pi_{cd}(k)D_{db}(k), \label{b3}
\end{equation}

\noindent
where $D_{ab}$ and $\Pi_{cd}$ are expressed in terms of 5$\times$5 matrices
 taking into account the mixture of 
 $\sigma$ and $\omega$ meson in nuclear matter.
 Meson propagator $D_{ab}$ is obtained by solving the Dyson equation 
 
\begin{equation}
 D_{ab}(k)=\left(\begin{array}{cccc}
             D_{\mu\nu} & D_{5\nu} \\
             D_{\mu 5} & D_{s} 
           \end{array}\right) ,\label{c1}
\end{equation}
\begin{displaymath}
  D_{\mu\nu}=\delta_{\mu\nu}D_l(k)+{k_\mu k_\nu \over k_\rho^2}
  \biggr \{ D_0(k)\biggr (1+{k_\lambda^2 \over m_v^2 }\biggr )-D_l(k)
  \biggr \} 
\end{displaymath}

\begin{equation}
   + \biggr (\delta_{ij}-{k_i k_j \over k^2}\biggr )(D_t(k)-D_l(k))\delta_{\mu i}
   \delta_{\nu j}, \label{c2}              
\end{equation}

\begin{equation}
 D_{5\nu}(k)=D_{\mu 5}(k)=\left\{\begin{array}{ll}
             i{k_i k_4 \over k_\rho^2}D_m(k) &  for \qquad \mu=i=1\sim 3 \\
             -iD_m(k) & for \qquad \mu=4 
            \end{array} \right. ,\label{c3}
\end{equation}

\noindent
where $D_0$ denotes the non-interacting meson propagator. The subscripts
 $s, l, t$
 and $m$ of $D(k)$ denote the component of $\sigma$ meson, the longitudinal
  and transverse component of $\omega$ meson and the component of mixture
  of $\sigma$ and $\omega$ mesons, respectively.
 Eqs.(\ref{b2}) and (\ref{b3}) are solved self-consistently.  
  
  The total energy density is given in six parts as follows,
  
  \begin{equation}
  \epsilon =
  \epsilon_{B}
  +\epsilon_{C,\sigma}
  +\epsilon_{C,\omega}
  +\epsilon_{SD,\sigma}
  +\epsilon_{SD,\omega}
  +\epsilon_{SD,m},  \label{c4}
  \end{equation}
  
  \noindent 
  where subscripts $B,C$ and $SD$ denote the baryon part, the 
  classical part and the quantum part, respectively. 
  The details of the energy densities of these components   are as follows,
  
  \begin{equation}
  \epsilon_{B}={\lambda \over \pi^2}\int_{0}^{k_F}q^2 dqE_q,
  \label{c5}
  \end{equation}
  
  \begin{equation}
  \epsilon_{C,\sigma}={1 \over 2}\biggr({g_s \over m_s }\biggr )^2\rho_s^2,~~~~
  \rho_{s}={\lambda \over \pi^2}\int_{0}^{k_F}k^2 dk{M_k^* \over E_k^*},
  \label{c6}
  \end{equation}
  
  \begin{equation}
  \epsilon_{C,\omega}=-{1 \over 2}\biggr({g_v \over m_v }\biggr )^2\rho_B^2,~~~~  \rho_{B}={\lambda \over 3\pi^2}k_F^3,
  \label{c7}
  \end{equation}
  
  \begin{displaymath}
  \epsilon_{SD,\sigma}={\lambda g_s^2 \over 16 \pi^4}
  \int_{0}^{k_F}{q^2 dq \over E_q^*}
  \int_{0}^{k_F}{k^2 dk \over E_k^*}
  \int_{-1}^{1}dx \{ 1-2(E_k -E_q)^2D_s^0(R)\} 
  \end{displaymath}
  
  \begin{equation}
  \times (M_k^* M_q^*-q^* k^* x +E_q^* E_k^*)D_s(R),
  \label{c8}
  \end{equation}
  
  \begin{displaymath}
  \epsilon_{SD,\omega}=-{\lambda g_v^2 \over 16 \pi^4}
  \int_{0}^{k_F}{q^2 dq \over E_q^*}
  \int_{0}^{k_F}{k^2 dk \over E_k^*}
  \int_{-1}^{1}dx \{ 1-2(E_k -E_q)^2D_v^0(R)\}
  \end{displaymath}
 
  \begin{displaymath}
 \times \biggr (4(M_k^* M_q^*+2k_\mu^* q_\nu^*)D_l(R)
           +\biggr\{ \biggr (4-{R_\mu^2 \over R^2 }\biggr )M_q^* M_k^* 
  \end{displaymath}
  
  \begin{equation}
  +\biggr\{ \biggr (2-{R_\mu^2 \over R^2 }\biggr )q^* k^* x
       - \biggr (2+{R_\mu^2 \over R^2 }\biggr )E_q^* E_k^* \biggr \}
         (D_t(k)-D_l(k))\biggr ),
  \label{c9}
  \end{equation}
  
  \begin{displaymath}
  \epsilon_{SD,M}={\lambda g_s g_v \over 8 \pi^4}
  \int_{0}^{k_F}{q^2 dq \over E_q^*}
  \int_{0}^{k_F}{k^2 dk \over E_k^*}
  \int_{-1}^{1}dx \{ 1-(E_k -E_q)^2(D_s^0(R)+D_v^0(R) \} 
  \end{displaymath}
  
  \begin{equation}
  \times {R_\mu^2 \over R^2}(E_q^* M_k^*+E_k^* M_q^*)D_m(R),
  \label{c10}
  \end{equation}
  
 \noindent
 where $E_k$ is a spectrum of the nucleon in the nuclear matter
 and $E_k^*=\sqrt{\vec k^{*2} +M_k^{*2}}$ and $M_k^*=M+\Sigma_s (k)$
 , 
 $R_\mu =k_\mu-q_\mu$ and $\lambda$ denotes the degeneracy, $\lambda =2$ for nuclear matter and $\lambda =1$ for neutron matter. 
 If we neglect the quantum parts of Eqs. (\ref{c8}) $\sim$ (\ref{c10}), the expressions correspond to the MFT. Similarly if one replace the full meson 
 propagator $D$ with the free one $D^0$ and neglect the mixture $D_m$, one obtain the HF .
 
 We neglect the other nonstrange mesons ($\pi ,\eta ,\delta$ , and $\rho$)
 because the contributions of these mesons for quark condensate
 are much smaller than the contributions of $\sigma$ and $\omega$ mesons 
 \cite{li} \cite{brockmann}.

In Fig.1, we show the quark condensate in nuclear matter calculated
 by some methods. The dotted line denotes the results in 
 the leading order approximation.
 The solid line, dashed line, and dot-dashed line denote the results based on 
 the BNSD , the MFT and the HF , 
 respectively. The scalar and vector coupling constants are chosen
 to satisfy the boundary condition at the normal density of nuclear matter
  ($\rho_0=$0.17$fm^{-3}$) for each methods.  
  In the BNSD, the HF and the MFT , 
  the quark condensate tends to increase around
  1.5 times of the normal density, 
  contrary to expectations based on
  ideas of chiral symmetry restoration. 
  Comparing the result of
  the BNSD with the one of the HF, 
  we notice the higher order ring correlation reduces 
  the quark condensate. 
  
  The detail of this reduction of quark condensate 
  is shown in Fig.2.
  In this figure we show the density dependence of derivative 
  of the energy per nucleon with 
  respect to the nucleon mass, the $\sigma$ meson mass and the $\omega$
   meson mass. The solid line and dot-dashed 
  line denote the results of  the BNSD and the HF, respectively. 
  The contribution
  from the $\sigma$  meson in the BNSD is larger than the one in the HF 
  by the correction of the higher order diagrams .
  Similarly the contribution
  from the $\omega$  meson in the BNSD is smaller than the one in the HF 
  by the correction of 
  the higher order diagrams.
  While the contribution from the nucleon 
  in the BNSD is nearly equal to the one in the HF.
  The higher order contribution from the nucleon mass derivative 
  is hardly   noticeable. 
  The contributions from the nucleon, the $\sigma$ meson and the 
  $\omega$ meson 
    cancel out one another. 
    The net contribution in the cancellation reduces the quark 
    condensate at the vacuum value.  
    Since the net contribution in the BNSD is larger than the one in the HF ,
    the quark condensate in the BNSD is reduced more than the one in the HF. 
   
 In summary, 
 we have calculated the quark condensate in nuclear matter using
 the BNSD and found that the quark condensate at the 
 normal density is reduced 
 about 35$\%$ compared to the one at zero density 
  in the BNSD as same as in the MFT and the HF (see Fig.1). 
 The difference of these three models 
 appears at high densities. 
 The higher order contribution of ring diagrams
  dose not affect the quark condensate at low densities. 
  But the contribution is important at high densities.
 
The authors gratefully acknowledge the computing time granted by 
 Research Center
 for Nuclear Physics (RCNP) .

\vfill\eject

\vfill\eject

\centerline{Figure caption}

Fig.1~~:~~The quark condensate in nuclear matter.   
  The dotted line denotes the results in the leading order approximation.
 The solid line, dashed line, and dot dash line denote the results based on 
 the BNSD, the MFT and the HF , 
 respectively. 
 
 \medskip
 
 Fig.2~~:~~Derivatives of the energy-per-nucleon ( $\partial (E/A) / \partial
  m_i$ ) in nuclear matter.
 The solid line  and dot dash line denote
 the results based on the BNSD and  the HF.

\end{document}